\documentclass[12pt]{iopart}

\usepackage{iopams}
\usepackage{color}
\begin{document}

\title[On the generalized continuity equation]{On the generalized continuity equation}

\author{Arbab I. Arbab$^{1}$ and Hisham. M. Widatallah$^{2}$}

\address{$^{1}$Department of Physics, Faculty of Science, University of Khartoum, P.O. Box 321, Khartoum
11115, Sudan, \\
$^{2}$Department of Physics, College  of Science, Sultan Qaboos University, P.O. Box 36 Al-Khod, Muscat, 123, Oman}
\ead{aiarbab@uofk.edu, hishamw@squ.edu.om}

\begin{abstract}
A generalized continuity equation extending the ordinary continuity equation has been found using quanternions. It is shown to be compatible with  Dirac, Schrodinger, Klein-Gordon and diffusion equations. This generalized equation is  Lorentz invariant. The  transport properties of electrons are found to be governed by Schrodinger-like equation and not by the diffusion equation.
\end{abstract}

\maketitle

\section{Introduction}
The continuity equation governs the conservation of mass/charge/ probability of any closed system. This equation involves the spatial distribution of the flux density that is related to the temporal variation of the particle density (charge/mass). Ordinarily, this equation is derived from  the equation of motion. The motion of any continuous charge/ mass distribution can be thought of a continuum (field or fluid). The continuity equation guarantees that no loss or gain of such quantities. This equation provides us with information about the system. The  information is carried from one point to another by a particle (field) wave. We therefore, trust that such an equation should exhibit this property. For electromagnetic field, the information about the motion of the charges is carried away by photons. Because electrons have particle-wave nature, the flow of  electrons is not like the flow of classical objects. The information about the movement of electrons is carried by the wave-particle nature that electrons have.  The current and charge/mass densities are unified if we formulate our basic equations governing a particular system in terms of Quaternion. We have recently, shown that Maxwell equations can be written as a single quaternionic equation [1]. Moreover,  we have found that the quaternionic Lorentz force led to the Biot-Savart law.  The magnetic field produced by the charged particles of the medium is found to be always perpendicular to the particle direction of motion.
This magnetic field produces a longitudinal wave called \emph{electroscalar} wave propagates with speed of light in vacuum.   Moreover, the electromagnetic field travels with speed of light in vacuum is the presence of a medium if the current and charge densities satisfy the generalized continuity equation.

We aim in this work at investigating a generalized equation expressing a wave-like nature for the charge and current densities,  using Quaternions and seek a wave-like continuity equation. Such a study will yield  the ordinary continuity equation and some other additional equations. This system of equations will reveal the wave nature of the current and charge(mass/probability) densities. We call the set of these equations the generalized continuity equation (GCE).   Maxwell, Schrodinger, Klein-Gordon and diffusion equations are shown to be compatible with this GCE.
Using the GCE, we  have shown that Ohm's law is equivalent to a Schrodinger-like equation. Hence, the electrical properties of metals are consequences of the  wave-particle behavior of electrons and not merely due to the drift of electrons.
\section{Continuity Equation}

\markright{Arbab I. Arbab On The New Gauge Transformations Of Maxwell's Equations}

The flow of any continuous medium is governed by the continuity equation. We would like here to write the continuity equation in terms of quaternions. The multiplication rule for the two quaternion, $\widetilde{A}=(a_0\,, \vec{A})$ and $\widetilde{B}=(b_0\,, \vec{B})$ is given by [2]
\begin{equation}
\widetilde{A}\widetilde{B}=\left(a_0b_0-\vec{A}\cdot\vec{B}\,\,, a_0\vec{B}+\vec{A}b_0+\vec{A}\times\vec{B}\right).
\end{equation}
Accordingly, one can write the quaternion continuity equation in the form
\begin{equation}
\widetilde{\nabla}\widetilde{J}=\left[-\,\left(\vec{\nabla}\cdot
\vec{J}+\frac{\partial \rho}{\partial t}\right) \,,\,
\frac{i}{c}\,\left(\frac{\partial \vec{J}}{\partial
t}+\vec{\nabla}\rho\, c^2\right)+\vec{\nabla}\times
\vec{J}\,\right]=0\,,
\end{equation}
where
\begin{equation}
\widetilde{\nabla}=\left(\frac{i}{c}\frac{\partial}{\partial
t}\, , \vec{\nabla}\right)\,,\qquad \widetilde{J}=\left(i\rho c\, ,
\vec{J}\right)\,.
\end{equation}
This implies that
\begin{equation}
\vec{\nabla}\cdot \vec{J}+\frac{\partial \rho}{\partial
t}=0\,,
\end{equation}
\begin{equation}
\vec{\nabla}\rho+\frac{1}{c^2}\frac{\partial
\vec{J}}{\partial t}=0\,,
\end{equation}
and
\begin{equation}
\vec{\nabla}\times
\vec{J}=0\,.
\end{equation}
We call Eqs.(3) - (5) the \emph{generalized continuity equation} (\textcolor[rgb]{1.00,0.00,0.00}{GCE}). Equation (6) states the current density $\vec{J}$ is irrotational.

In a covariant form, Eqs.(4) - (6) read
\begin{equation}
\partial_\mu J^\mu=0\,,\qquad N_{\mu\nu}\equiv\partial_\mu J_\nu-\partial_\nu J_\mu=0\,.
\end{equation}
Notice that the  tensor $N_{\mu\nu}$ is an antisymmetric tensor. It is evident from Eq.(7) that Eqs.(4) - (6) are Lorentz invariant.
Now differentiate Eq.(4) partially with respect to time and use Eq.(5), we obtain
\begin{equation}
\frac{1}{c^2}\frac{\partial^2\rho}{\partial t^2}-\nabla\,^2\rho=0\,,
\end{equation}
Similarly,  take the divergence of  Eq.(5) and use Eq.(4), we obtain
\begin{equation}
\frac{1}{c^2}\frac{\partial^2\vec{J}}{\partial
t^2}-\nabla\,^2\vec{J}=0\,,
\end{equation}
where $\rho=\rho(r,t)$ and $\vec{J}=\vec{J}(\vec{r},t)$. Therefore, both the current density and charge density satisfy the solution of a wave equation propagating with  speed of light.
We remark that the \textcolor[rgb]{1.00,0.00,0.00}{GCE} is applicable to any flow whether created by charged particles or neutral ones.

\markright{Arbab I. Arbab On The New Gauge Transformations of Maxwell's Equations}
\section{Maxwell's Equations}
\markright{Arbab I. Arbab On The New Gauge Transformations of Maxwell's Equations}
Maxwell's equations are given by [3]
\begin{equation}
\hspace{-2.5cm}\vec{\nabla}\cdot \vec{E}=\frac{\rho}{\varepsilon_0}\,\,,\qquad \vec{\nabla}\cdot \vec{B}=0\,,
\qquad
\vec{\nabla}\times \vec{E}+\frac{\partial\vec{B}}{\partial t}=0\,, \qquad \vec{\nabla}\times \vec{B}-\frac{1}{c^2}\frac{\partial\vec{E}}{\partial t}=\mu_0\vec{J}\,.
\end{equation}
Now differentiate Faraday's equation partially with respect to time and  employ Ampere's equation and the vector identity $\vec{\nabla}\times\left(\vec{\nabla}\times\vec{B}\right)=\vec{\nabla}(\vec{\nabla}\cdot\vec{B})-\nabla^2\vec{B}$\,  to get

\begin{equation}
\frac{1}{c^2}\frac{\partial^2\vec{B}}{\partial
t^2}-\nabla\,^2\vec{B}=\mu_0\left(\vec{\nabla}\times \vec{J}\right)\,.
\end{equation}
Similarly, differentiating Ampere's equation partially with respect to time and employing Gauss and Faraday equations yields
\begin{equation}
\frac{1}{c^2}\frac{\partial^2\vec{E}}{\partial
t^2}-\nabla^2\vec{E}=-\frac{1}{\varepsilon_0}\left(\vec{\nabla}\rho +\frac{1}{c^2}\frac{\partial
\vec{J}}{\partial t}\right)\,,
\end{equation}
According to the \textcolor[rgb]{1.00,0.00,0.00}{GCE} the left hand sides of Eqs.(11) and (12) vanish.  This entitles us to say that the electromagnetic field travels with speed of light in free space (when $\rho=0,\vec{J}=0)$ and in presence of charge and current; as far as the current and charge densities satisfy the \textcolor[rgb]{1.00,0.00,0.00}{GCE}.

Now consider a conducting media defined by the conduction current density (Ohm's law)
\begin{equation}
\vec{J}=\sigma\,\vec{E}\,.
\end{equation}
Taking the divergence of Eq.(13) and using the Gauss law, one obtains
\begin{equation}
\vec{\nabla}\cdot\vec{J}=\sigma\,\vec{\nabla}\cdot\vec{E}=\sigma\frac{\rho}{\varepsilon_0}\,.
\end{equation}
Differentiating Eq.(14) partially with respect to time and using Eq.(5) yields
\begin{equation}
\nabla^2\rho=-\frac{\sigma}{c^2\varepsilon_0}\frac{\partial\rho}{\partial t}\,,\qquad \frac{\partial\rho}{\partial t}=-D_c\nabla^2\rho\,,\qquad D_c=\frac{c^2\varepsilon_0}{\sigma}=\frac{1}{\mu_0\sigma}\,.
\end{equation}
 Equation (15) is a Schrodinger-like wave equation describing the motion of free electrons. A metal  can be viewed  as a gas of free electrons (Fermions). The thermal properties of these electrons are determine by Fermi-Dirac statistics.  Thus, Ohm's law satisfies Schrodinger equation. This is a very interesting result, since the electron motion is governed by Schrodinger equation owing to the wave-particle nature of the electrons. Thus, electrons move in a metal as a material wave (de Broglie wave).

 The Drude has proposed a model of electrical conduction of electrons to explain the transport properties of electrons in materials (especially metals) [4]. The model, which is an application of kinetic theory, assumes that the microscopic behavior of electrons in a solid may be treated classically and looks much like a pinball machine, with a sea of constantly jittering electrons bouncing and re-bouncing off heavier, relatively immobile positive ions. He has further  shown that the Ohm's law is true, where he related the current density to the electric field by a linear similar formula. He also connected the current as a response to a time-dependent electric field by a complex conductivity.
This simple classical Drude model provides a very good explanation of DC and AC conductivity in metals, the Hall effect, and thermal conductivity (due to electrons) in metals. The failure of Drude model to explain specific heat of metal is connected with its disregard to the wave nature of electrons.
Hence, the motion of electrons in metals in not due to diffusion of electrons, the electrons is transported quantum mechanically.  Hence, a quantum Ohm's law should be considered instead in which the current density $\vec{J}$ is obtained from Schrodinger theory of a free electron (Fermi gas).
Thus,  quantum mechanically, an electron is viewed as a wave traveling through a medium. When the wavelength of the electrons is larger than the crystal spacing, the electrons will propagate freely throughout the metal without collision, therefore their scattering  result only from the imperfections in the crystal lattice of the metal.

 In Schrodinger terminology the diffusion constant $D_c=\frac{\hbar}{2m\,i}$, which shows that the diffusion constant is complex. This would imply that the conductivity is complex. In Schrodinger paradigm, the current density and charge density are probabilistic quantities determined by the wave function of the electron ($\psi$) as [5]
\begin{equation}
 \rho=\psi^*\psi\,, \qquad \vec{J}=\frac{\hbar}{2mi}(\psi^*\nabla\psi-\psi\nabla\psi^*)
 \end{equation}
 The current density and the charge here are related by the generalized continuity equation. Therefore, the Ohm's current is a probabilistic wave-like current.

 It is thought that electrons move in a metal by drifting, but we have seen here the electrical transport properties of metals are propagated by a a material wave rather than drifting of electrons.

 Now, apply Eq.(4) in Eq.(13) and employ Gauss law to get
\begin{equation}
\frac{\partial\rho}{\partial t}=-\frac{\sigma}{\varepsilon_0}\,\rho\,.
\end{equation}
Let us write the charge density is a separable form
\begin{equation}
\rho(r,t)\equiv \rho(r)\rho(t)\,.
\end{equation}
The time dependence of the charge density $\rho(r,t)$ is obtained from Eq.(17) as
\begin{equation}
\rho(t)=\rho_0\exp(-\frac{\sigma}{\varepsilon_0})\,t\,,\qquad \rho_0=\rm const.
\end{equation}
The constant $\tau\equiv\frac{\sigma}{\varepsilon_0}$ is know as the relaxation time, and it is a measure of how fast a conducting medium reaches  electrostatic equilibrium.
The spatial dependence of the charge density is obtain from substituting Eq.(17) in Eq.(15). This yields the equation
\begin{equation}
\nabla^2\rho (r)-\frac{1}{\lambda_c^2}\rho(r)=0\,,\qquad \lambda_c=\frac{\varepsilon_0c}{\sigma}=\frac{1}{\mu_0c\,\sigma}=\frac{D_c}{c}\,.
\end{equation}
The solution of Eq.(20) is given by
\begin{equation}
\rho(r)=A\exp(-\frac{r}{\lambda_c})\,,\qquad A=\rm const.
\end{equation}
Therefore, the charge density distribution (space-time) is given by
\begin{equation}
\rho(r,t)=Be^{-\frac{1}{\lambda_c}(\,r+c\,t)}+Fe^{-\frac{1}{\lambda_c}(-\,r+c\,t)}\,,\qquad B\,, F=\rm const
\end{equation}
This shows that the charges density decays in space and time. However, if $\omega_c$ is complex, we will have an oscillatory charge density solution. This is true if we consider  $\sigma$ to be complex.
The spreading of wavepackets in quantum mechanics is directly related to the spreading of probability densities in diffusion (see sec. 4 below).

Now take the gradient of Eq.(14) and employ Eq.(5) and the vector identity $\nabla\times(\vec{\nabla}\times \vec{J})=\nabla(\vec{\nabla}\cdot \vec{J})-\nabla^2\vec{J}$\, [6] we obtain
\begin{equation}
\frac{\partial \vec{J}}{\partial t}=-D_c\nabla^2\vec{J}\,.
\end{equation}
This equation shows that $\vec{J}$ is governed by Schrodinger-like equation. Hence, both the current density and charged density are governed by a Schrodinger-like equation.

Now applying Eq.(13) to Eq.(5) and using Eq.(10), yield
\begin{equation}
\frac{\partial \vec{E}}{\partial t}=-D_c\nabla^2\vec{E}\,,
\end{equation}
where we have used the vector identity  $\nabla\times(\vec{\nabla}\times \vec{E})=\nabla(\vec{\nabla}\cdot \vec{E})-\nabla^2\vec{E}$ and the fact that $\vec{\nabla}\times \vec{E}=0$. This equation can also be obtained directly from Eq.(23) by dividing both sides of Eq.(23) by $\sigma$ and using Eq.(13).

Equation (20) may define a limiting conductivity of a material ($\sigma_0$), when we equate $\lambda_c$ to Compton wavelength of the electron of mass $m_e$. In this case one finds
\begin{equation}
\sigma_0=\frac{m_e}{\mu_0h}\,.
\end{equation}
This amounts to a value of $\sigma_0=1.09\times10^{9}\Omega^{-1}\, m^{-1}$. This can be compared with the best conductivity of metal (Silver) which is $6.3\times 10\,^{7}\Omega^{-1}\, m^{-1}$. Thus, the maximum possible conductivity is set by quantum mechanics and governed by Eq.(25). Because of this the conductivity at zero Kelvin is never infinite but should be limited to $\sigma_0$.
\subsection{Fluid nature of electromagnetic field}
We have recently found that the magnetic field created by the charged particle to be given by [1]
\begin{equation}
\vec{B}_m=\frac{\vec{v}}{c^2}\times\vec{E}\,.
\end{equation}
In magnetohydrodynamics this field is govern by the magnetohydrodynamic equation
\begin{equation}
\frac{\partial \vec{B}}{\partial t}=\vec{v}\times(\nabla\times \vec{B})\,.
\end{equation}
For any other fluid one has in general
\begin{equation}
\frac{\partial \vec{\omega}}{\partial t}=\vec{v}\times(\nabla\times \vec{\omega})\,,
\end{equation}
where $\omega=\vec{\nabla}\times\vec{v}$\, is the vorticity of the fluid. This shows that the vorticity of a fluid is an integral part to the fluid flow. When an electromagnetic field encounters a charged particle, the charged particle induces a vorticity in the field (or fluid). This equally occurs in fluids when a fluid encounters an obstacle. I have recently shown that Eq.(6), i.e., $\vec{\nabla}\times \vec{J}=0$,  implies that $\vec{\omega}=\frac{\vec{v}}{c^2}\times (-\frac{\partial \vec{v}}{\partial t})$ [7]. Owing to the magnetic field in Eq.(24), we have shown recently that this field gives rise to a longitudinal wave (electroscalar wave) traveling at speed of light in vacuum besides the electromagnetic fields [1].

Hence, it is evident that there is a genuine one-to-one  correspondence between electrodynamics and hydrodynamics, i.e., they are intimately related. This would immediately imply that the electromagnetic fields propagation mimics the  fluids flow. The electric field resembles the local acceleration and the magnetic fields resembles the vorticity of the moving fluid.

Using Eq.(13) in Eq.(24),  the magnetic field (and vorticity) produced by the electrons in a metal vanishes, since  $\vec{B}_m=\frac{\vec{v}\times \vec{J}}{c^2\sigma}=0$, where $\vec{J}=\rho \,\vec{v}.$ Hence, the Lorentz force on conduction electrons is only electric. Hence, no magnetic field inside the conductor can be created from the motion of electrons. Therefore, even if we write Ohm's law in the general form $\vec{J}=\sigma(\vec{E}+\vec{v}\times\vec{B})$, we would have obtained the form in Eq.(13) too.

\markright{Arbab I. Arbab On The New Gauge Transformations of Maxwell's Equations}
\section{Diffusion Equation}
 Diffusion is a transport phenomena resulting from random molecular motion of molecules from a region of higher concentration to a lower concentration. The result of diffusion is a gradual mixing of material.  Diffusion is of fundamental importance in  physics, chemistry, and biology.

\markright{Arbab I. Arbab On The New Gauge Transformations of Maxwell's Equations}
The diffusion equation is given by [8]
\begin{equation}
\vec{J}=-D\vec{\nabla}\rho\,,
\end{equation}
where  $D$ is the diffusion constant. This is known as Fick's law.
Taking the divergence of Eq.(27) and using Eq.(4), one finds
\begin{equation}
\frac{\partial \rho}{\partial t}=D\nabla^2 \rho\,.
\end{equation}
This shows that the density $\rho$ satisfies the Diffusion equation. The normalized solution of the Eq.(28) is
\begin{equation}
 \rho (x,t)= \frac{1}{\sqrt{4\pi D\,t}} \exp\,(-\frac{x^2}{4Dt})\,,
\end{equation}
which when applied in Eq.(27) yields the current density $\vec{J}$. It is obvious that the  current in Eq.(27) satisfies the \textcolor[rgb]{1.00,0.00,0.00}{GCE}, viz., Eq.(4) - (6).
Differentiation Eq.(27) partially with respect to time and employing Eqs.(4) - (5) and the vector identity, $\vec{\nabla}(\vec{\nabla}\cdot\vec{J})=\nabla^2\vec{J}+\vec{\nabla}\times(\vec{\nabla}\times \vec{J})$\,[6] one gets
\begin{equation}
\frac{\partial \vec{J}}{\partial t}=D\nabla^2 \vec{J}\,.
\end{equation}
Hence, both $\rho$ and $\vec{J}$ satisfy the diffusion equation.
\section{Dirac Equation}
We would like here to show that the generalized continuity equations, viz., Eq.(7) are compatible with Dirac equation. To this aim, we apply the current density 4-vector according to Dirac formalism, i.e., $J_\mu=\overline{\psi}\gamma_\mu \psi$  in  Eq.(7). This yields [5]
\begin{equation}\label{3}
\partial^\mu J^\nu-\partial^\nu J^\mu =(\partial^\mu\overline{\psi})\gamma^\nu \psi+\overline{\psi}\gamma^\nu \partial^\mu \psi -
(\partial^\nu\overline{\psi})\gamma^\mu \psi-\overline{\psi}(\gamma^\mu \partial^\nu \psi)=0.
\end{equation}
The first term in the above equation can be written as
\begin{equation}\label{3}
(\partial^\mu\overline{\psi})\gamma^\nu \psi=(\partial^\mu\psi^+)\gamma^0\gamma^\nu \psi=(\partial^\mu\psi^+)\gamma^{\nu+}\gamma^0 \psi=(\gamma^\nu\partial^\mu\psi)^+\gamma^0 \psi.
\end{equation}
Hence,
\begin{equation}
(\partial^\mu\overline{\psi})\gamma^\nu \psi=(\overline{\psi}\gamma^\nu \partial^\mu \psi)^+.
\end{equation}
Similarly,
\begin{equation}
(\partial^\nu\overline{\psi})\gamma^\mu \psi=(\overline{\psi}\gamma^\mu \partial^\nu \psi)^+.
\end{equation}
Substituting  these terms in Eq.(31) completes the proof, since for a plane wave $\partial^\mu\psi=ik^\mu\psi$.
\section{Klein-Gordon Equation}
It is the equation of motion of scalar particles with integral spin. For such a system  the current density 4-vector is given by [5]
\begin{equation}\label{3}
J_\mu=\frac{i\hbar}{2m}\left(\phi^*\partial_\mu\phi-(\partial_\mu\phi^*)\phi\right),
\end{equation}
so that the generalized continuity equation  becomes
\begin{equation}\label{3}
\hspace{-2cm}\partial_\mu J_\nu-\partial_\nu J_\mu =\frac{i\hbar}{2m}\left[\partial_\mu(\phi^*\partial_\nu\phi-(\partial_\nu\phi^*)\phi)-\partial_\nu(\phi^*\partial_\mu\phi-(\partial_\mu\phi^*)\phi)\right]=0,
\end{equation}
and hence, Klein-Gordon equations is compatible with our \textcolor[rgb]{1.00,0.00,0.00}{GCE} too.
\section{Schrodinger Equation}
It the equation of motion governing the motion of non-relativistic particles, eg. electrons.
The current density and probability density in Schrodinger formalism are given by [5]
\begin{equation}\label{3}
\rho=\psi^*\psi,\qquad \vec{J}=\frac{\hbar}{2mi}(\psi^*\nabla\psi-(\nabla\psi^*)\psi).
\end{equation}
Applying Eq.(6), one gets
\begin{equation}
\vec{\nabla}\times \vec{J}=\frac{\hbar}{2\,mi}\left[(\vec{\nabla}\times(\psi^*\nabla\psi)-\vec{\nabla}\times(\nabla\psi^*)\psi)\right].
\end{equation}
Using the two vector identities $\vec{\nabla}\times(f\vec{A})=f(\vec{\nabla}\times \vec{A})-\vec{A}\times(\vec{\nabla}f)$ and $\vec{\nabla}\times(\vec{\nabla} f)=0$, Eq.(38) vanishes, and hence  one of the  \textcolor[rgb]{1.00,0.00,0.00}{GCE} is satisfied. We now apply  Eq.(5) in Eq.(37) to get
\begin{equation}
\vec{\nabla}(\rho c^2)+\frac{\partial \vec{J}}{\partial t}=\vec{\nabla}(\psi^*\psi)+\frac{\hbar}{2\,mi}\frac{\partial}{\partial t}\left[\,\psi^*\nabla\psi-(\nabla\psi^*)\,\psi\,\right]=0.
\end{equation}
Upon using the Schrodinger equation,
\begin{equation}
H\psi=i\hbar\frac{\partial \psi}{\partial t}\,, \qquad \psi^*H=-i\hbar\frac{\partial \psi^*}{\partial t}\,
\end{equation}
Eq.(39) vanishes for a plane wave equation, i.e., $\psi\,(r,t)=A\exp\,i(\vec{k}\cdot \vec{r}-\omega\,t)\,, A={\rm const.}$. Therefore, the \textcolor[rgb]{1.00,0.00,0.00}{GCE} is satisfied by Dirac, Klein-Gordon and Schrodinger equations. This implies that these \textcolor[rgb]{1.00,0.00,0.00}{GCE} is indeed fundamental in formulating any field theoretic models involving the motion of particles or fluids. This equation will have immense consequences when applied to
the theory of electrodynamics. Such an application will lead to  better formulations and understanding of the astrophysical phenomena pertaining to the evolution of dense objects.
\section{Concluding Remarks}
 We have derived in this paper the  generalized continuity equations and showed how it is applicable to the particle flow.   We have  shown that the the generalized continuity equation is lorentz invariant. We have then shown that the basic equations of motion are compatible with the generalized continuity equation. The diffusion equation is in agreement with the \textcolor[rgb]{1.00,0.00,0.00}{GCE}. Moreover,  the current and density in Dirac,  Klein-Gordon and Schrodinger formalisms are compatible with the \textcolor[rgb]{1.00,0.00,0.00}{GCE} too. The classical Ohm's law is found to be compatible with Schrodinger-like equation. Transport properties of metals are shown to be propagated by a material wave rather than drifting and diffusion.
\section*{Acknowledgments} One of us (AIA) would like to thank the university of Khartoum for financial support of this work and the Sultan Qaboos university for hospitality where this work is carried out.
\section*{References}
$[1]$ Arbab, A. I., and Satti, Z. A.,  \emph{Progress in Physics}, 2, \textbf{8} (2009).\\
$[2]$ Tait, Peter Guthrie,  \emph{An elementary treatise on quaternions}, 2nd  ed., Cambridge  University Press (1873);
Kelland, P. and Tait, P. G., \emph{Introduction to Quaternions}, 3rd ed. London: Macmillan, (1904).\\
$[3]$ Jackson, J.D.,   \emph{Classical Electrodynamics}, New York, John Wiley, 2nd ed. (1975).\\
$[4]$ Drude, Paul,  \emph{Annalen der Physik} \textbf{306}, (3), 566 (1900).\\
$[5]$ Bjorken, J.D. and Drell, S.D.,  \emph{Relativistic Quantum Mechanics}, McGraw-Hill Book Company, (1964).\\
$[6]$ Lawden, D.F., \emph{Tensor Calculus and Relativity}, Methuen, London, (1968).\\
$[7]$ Arbab, A. I.,  \emph{\tt On the analogy between the electrodynamics and hydrodynamics using quaternions},  to appear in the 14th International Conference on Modelling Fluid Flow (CMFF'09), Budapest, Hungary, 9-12 September (2009).\\
$[8]$ Fick A., \emph{Ann. Physik, Leipzig}, \textbf{170}, 59, (1855).\\
\end{document}